\documentclass[prd,superscriptaddress,onecolumn,showpacs,showkeys]{revtex4}
\usepackage{eurosym}
\usepackage{amsfonts}
\usepackage{array}
\usepackage{comment}
\usepackage{amsthm}
\usepackage{bm}
\usepackage{changes}
\usepackage{palatino}
\usepackage[colorinlistoftodos]{todonotes}
\usepackage{mathpazo}
\usepackage{amssymb}
\usepackage{eurosym}
\usepackage{amsmath}
\usepackage{epsfig}
\usepackage{graphics}
\usepackage{color}
\usepackage{graphicx}

\def\be{\begin{equation}}
\def\ee{\end{equation}}
\def\beq{\begin{eqnarray}}
\def\eeq{\end{eqnarray}}

\def\bes{\begin{eqnarray}}
\def\ees{\end{eqnarray}}

\usepackage{hyperref}
\hypersetup{
    colorlinks=true,
    linkcolor=blue,
    filecolor=magenta, citecolor=blue,    urlcolor=blue,
}

\begin{document}

\title{Charged fermions tunneling from stationary axially symmetric black
holes with generalized uncertainty principle}
\author{Muhammad Rizwan} 
\email{m.rizwan@sns.nust.edu.pk}
\affiliation{Department of Computer Science, Faculty of Engineering \& Computer Sciences, National University of Modern Languages, H-9, Islamabad, Pakistan}
\affiliation{
School of Natural Sciences, National University of Sciences and 
Technology, H-12, Islamabad, Pakistan.}

\author{Muhammad Zubair Ali}
\email{mzam1@student.waiako.ac.nz}
\affiliation{Department of Mathematics and Statics, Faculty of Computational and Mathematical Sciences, University of Waikato Hamilton, 3216, New Zealand}

\author{Ali \"{O}vg\"{u}n}
\email{ali.ovgun@pucv.cl}
\homepage{http://www.aovgun.com}
\affiliation{Instituto de F\'{\i}sica, Pontificia Universidad Cat\'olica de
Valpara\'{\i}so, Casilla 4950, Valpara\'{\i}so, Chile.}
\affiliation{Physics Department, Arts and Sciences Faculty, Eastern Mediterranean University, Famagusta, North Cyprus via Mersin 10, Turkey.}

\begin{abstract}
In this paper, we study the tunneling of  charged fermions from the stationary axially symmetric black holes using the generalized uncertainty principle (GUP) via Wentzel, Kramers, and Brillouin (WKB) method. The emission rate of the charged fermions and corresponding modified Hawking temperature of Ker-Newman black hole, Einstein-Maxwell-Dilaton-Axion (EMDA) black hole, Kaluza-Klein dilaton black hole, and then, charged rotating black string  are obtained and we show that the corrected thermal spectrum is not purely thermal because of the minimal scale length which cause the black hole's remnant.

\end{abstract}


\date{\today}
\maketitle

\section{Introduction}

According to general relativity, black holes are so dense objects in the universe that not even light can escape.  The classical Einstein equations state that, black holes are disturbingly simple; their only properties are mass, electrical charge and angular momentum. The simplest solution of the Einstein equations in general spherically symmetric vacuum is a Schwarzschild solution, which depends only on a single parameter of mass. In addition,
the Reissner-Nordstr\"{o}m solution, has two parameters of electric charge and mass, which is the general spherically symmetric solution to
the Einstein-Maxwell (EM) equations. On the other hand, the axially symmetric solution of EM equations is Kerr-Newman solution which depends on three parameters; mass M, electric charge Q, and angular momentum J. Furthermore, one can use scalar fields together with string theory to solve mysteries of the dark matter and dark energy. The Einstein-Maxwell-dilaton-axion (EMDA) gravity is the low energy limit of the bosonic sector of the heterotic string theory \cite{1204.4319v2(26)}. EMDA gravity model is a  generalization of
EM gravity which contains the dilaton and the axion scalar fields. Moreover. black holes are interpreted microscopically in string theory as bound states of explicitly
specified constituents. For example: the original Kaluza-Klein theory in four dimensions, obtained by compactification
of five-dimensional pure gravity on a circle which contains the following fields:  a U(1) gauge field, a scalar field, and gravity \cite{bh1,bh2}. 

In 1974, Hawking predicted that black holes would release black body radiation, known as Hawking Radiation \cite{1010.6106v2(1),1010.6106v2(2)}. Moreover, this effect cause the information loss paradox, because of the thermal nature of the radiation. Nowadays, there have been many works on the derivation of the Hawking temperature using various techniques and methods \cite{1010.6106v2(3),1010.6106v2(4),1010.6106v2(5)}.
One of the famous method is the quantum tunneling method firstly used by Krauss-Wilczek-Parkih and also semiclassical method of tunneling using the Hamilton-Jacobi approach \cite{mKN,1312.3781v2(8i),aa1,aa2,aa3,aa4,aa5,aa6,aa7,aa8,aa9,aa10,aa11,aa12,aa13,aa14,za1,za2,za3,za4,za5,za6,za7,1312.3781v2(3a),1312.3781v2(3b),1312.3781v2(4a),1312.3781v2(4b),1307.0172v2(6),1312.3781v2(8a),1312.3781v2(8j),1312.3781v2(8k),1312.3781v2(8l),Javed:2018ufh,Javed:2018msn,Javed:2017cok,Javed:2017saf,Sharif:2012se,Sharif:2013nda,Sharif:2012xq,Sharif:2011gu,Sharif:2010pj,3DKN,Meitei:2018mgo,Singh:2017car,Singh:2017mqv,IbungochoubaSingh:2016jkk,IbungochoubaSingh:2016prd,Singh:2015zla,Meitei:2014oja,IbungochoubaSingh:2013jya,IbungochoubaSingh:2013kga,Ovgun:2017pvx}. Recently, the effect of the quantum gravity has been investigated using the generalized uncertainty principle (GUP) in different spacetimes, which indicate that the rate of the tunneling of particles deviates from pure thermality and satisfy the unitary theory. Furthermore, the researches on string theory, loop quantum gravity, double special relativity show that there is a possibility to existences of the minimal observable length which is the main ingredient of the GUP \cite{1312.3781v2(10),1312.3781v2(11),1312.3781v2(12),1312.3781v2(13),1312.3781v2(14),1312.3781v2(15)}. Briefly, this modification on the  uncertainty principle is as follows \cite{1312.3781v2(16)}:

\begin{equation*}
\left[ x_{i},p_{i}\right] =i\hbar \delta _{ij}\left[ 1+\beta p^{2}\right] ,
\end{equation*}%
where%
\begin{eqnarray*}
x_{i} &=&x_{0i},\text{ }p_{i}=p_{0i}\left( 1+\beta p^{2}\right) ,\text{ } \\
\text{ }p^{2} &=&p_{i}p^{i}\simeq -\hbar \left[ \partial _{i}\partial
^{i}-2\beta \hbar ^{2}\left( \partial _{j}\partial ^{j}\right) \left(
\partial ^{i}\partial _{i}\right) \right] .
\end{eqnarray*}%
Note that  $x_{0i}$ and $p_{0i}$ satisfy the canonical commutation relations $%
[x_{0i};p_{0j}]=$ $i\hbar \delta _{ij}$ and $\beta =\beta
_{0}l_{p}^{2}/\hbar ^{2},$ $\beta _{0}$ are dimensionless parameters with the Planck length $l_{p}$. Using these
commutation relations, the GUP can be written as \cite{1312.3781v2(16)}%
\begin{equation*}
\Delta x\Delta p\geq \frac{\hbar }{2}\left[ 1+\beta \left( \Delta p\right)
^{2}\right].
\end{equation*}%

In recent years there have been many publications including the effect of GUP \cite{aa15,aa16,aa17,aa18,aa19,aa20,aa21,1404.6375v1(34),1404.6375v1(38),1404.6375v1(39),1404.6375v1(40),1312.3781v2(31),1312.3781v2(33),1307.0172v2(25),1307.0172v21,1312.3781v2,1404.6375v1,1410.5075v1,1312.2075v1,rizwan}. Moreover, recently, tunneling of the uncharged particles from rotating black holes has been studied \cite{KN}, however, there is not completely agreement with the literature. The main aim of the paper is to obtain correct Hawking temperature using the tunneling of fermions from the stationary axially symmetric black holes. The idea is to get the quantum signature of the correlated Hawking quanta as a proof of the Hawking effect and to acquire the GUP effects on Hawking temperature, we study the general stationary axially symmetric black holes.

The paper is organized as follows: In Sec. II, we briefly review the method of tunneling using the charged fermions from the stationary axially symmetric black holes via GUP. In Sec. III, 
the modified Hawking temperature of Ker-Newman black hole is obtained. In Sec. IV, we study the temperature of  Einstein-Maxwell-Dilaton-Axion (EMDA) black hole. In Sec. V, we calculate the modified temperature of Kaluza-Klein dilaton black hole. Then, in Sec. VI, we obtain the effect of the
GUP on the charged rotating black string. In Sec. VII, we summarize our results.

\section{Modified temperature for general stationary axially symmetric black
hole}

In this section, we develop a general method to study charged fermions tunneling from stationary
axially symmetric black holes with GUP. We consider general $4-$dimensional
line element for stationary axially symmetric black holes and discuss
tunneling with general line element. Then we calculate the tunneling probability and we
give general formula for modified Hawking temperature. The general line
element and electromagnetic potential of axially symmetric black hole can be
written as%
\begin{equation}
ds^{2}=g_{tt}dt^{2}+g_{rr}dr^{2}+g_{\xi \xi }d\xi ^{2}+g_{\phi \phi }d\phi
^{2}+2g_{t\phi }dtd\phi, \label{1}
\end{equation}%
with%
\begin{equation}
A_{\mu }=A_{t}dt+A_{\phi }d\phi .  \label{2}
\end{equation}%
The line element can be transformed into the following form%
\begin{equation}
ds^{2}=\left( g_{tt}-\frac{g_{t\phi }^{2}}{g_{\phi \phi }}\right)
dt^{2}+g_{rr}dr^{2}+g_{\xi \xi }d\xi ^{2}+g_{\phi \phi }\left( d\phi +\frac{%
g_{t\phi }}{g_{\phi \phi }}dt\right) ^{2}.  \label{3}
\end{equation}%
The angular velocity of the black hole with line element (\ref{1}) is
defined as%
\begin{equation}
\Omega =-\frac{g_{t\phi }}{g_{\phi \phi }}.  \label{4}
\end{equation}%
It is a well known fact that the tunneling probability of particle is independent from the coordinates system, so
for simplicity of our calculation, we perform dragged coordinate
transformation. After transformation of coordinates as $d\phi =\Omega dt,$
the dragged metric can be written as%
\begin{eqnarray}
d\hat{s}^{2} &=&\left( g_{tt}-\frac{g_{t\phi }^{2}}{g_{\phi \phi }}\right)
dt^{2}+g_{rr}dr^{2}+g_{\xi \xi }d\xi ^{2},  \notag \\
d\hat{s}^{2} &\equiv &-Fdt^{2}+\frac{dr^{2}}{G}+H^{2}d\xi ^{2}.  \label{5}
\end{eqnarray}%
The transformed metric represents $3-$dimensional hyperspace in the $4-$%
dimensional spacetime. The corresponding electromagnetic vector potential is%
\begin{equation}
\hat{A}_{i}=\hat{A}_{t}dt=\left( A_{t}+\Omega A_{\phi }\right) dt.  \label{6}
\end{equation}%
Here quantities are three dimensional quantities. To discuss the
charged fermions tunneling with GUP, we use modified Dirac equation. The
modified Dirac equation for the fermions field of mass $m$ and charge $e$ can be
written as \cite{gupdirac}%
\begin{gather}
\left[ i\gamma ^{0}\partial _{0}+i\gamma ^{i}\partial _{i}\left( 1-\beta
m^{2}\right) +i\gamma ^{i}\beta \hbar ^{2}\left( \partial _{j}\partial
^{j}\right) \partial _{i}+\frac{m}{\hbar }\left( 1+\beta \hbar ^{2}\partial
_{j}\partial ^{j}-\beta m^{2}\right) \right.  \notag \\
\left. -\gamma ^{\mu }\frac{e}{\hbar }\hat{A}_{\mu }\left( 1+\beta \hbar
^{2}\partial _{j}\partial ^{j}-\beta m^{2}\right) +i\gamma ^{\mu }\Omega
_{\mu }\left( 1+\beta \hbar ^{2}\partial _{j}\partial ^{j}-\beta
m^{2}\right) \right] \Psi =0,  \label{MDE}
\end{gather}%
where $\Omega _{\mu }=\frac{1}{2}i\Gamma _{\mu }^{\nu \lambda }\Sigma _{\nu
\lambda },$ $\Sigma _{\mu \nu }=\frac{1}{4}i\left[ \gamma ^{\mu },\gamma
^{\nu }\right] $ and $\left\{ \gamma ^{\mu },\gamma ^{\nu }\right\} =2g^{\mu
\nu }I,$ with $i,$ $j=1,2$ and $\mu ,$ $\nu ,$ $\lambda =0,1,2.$ For
transformed metric (\ref{5}) gamma matrices $\gamma ^{\mu }$ can be
constructed as%
\begin{equation}
\gamma ^{t}=\frac{1}{\sqrt{F}}\left( 
\begin{array}{cc}
i & 0 \\ 
0 & -i%
\end{array}%
\right) ,\text{ }\gamma ^{r}=\sqrt{G}\left( 
\begin{array}{cc}
0 & \sigma ^{3} \\ 
\sigma ^{3} & 0%
\end{array}%
\right) ,\text{ }\gamma ^{\xi }=\frac{1}{H}\left( 
\begin{array}{cc}
0 & \sigma ^{1} \\ 
\sigma ^{1} & 0%
\end{array}%
\right) ,  \label{3.5}
\end{equation}%
where $\sigma ^{\prime }s$ are Pauli matrices. For fermions, there are two
states corresponding to spin up and spin down particles. The analysis for
both cases is same, so in this paper, we consider only the spin up
case. We assume the wave function as%
\begin{equation}
\Psi \left( t,r,\xi \right) =\left( 
\begin{array}{c}
A\left( t,r,\xi \right) \\ 
0 \\ 
B\left( t,r,\xi \right) \\ 
0%
\end{array}%
\right) \exp \left[ \frac{i}{\hbar }I\left( t,r,\xi \right) \right],
\label{3.6}
\end{equation}%
where $I$ is action of emitted fermion. On substitution of the wave function (%
\ref{3.6}) and the gamma matrices (\ref{3.5}) into the modified Dirac
equation (\ref{MDE}) we get the action form of modified Dirac equation%
\begin{eqnarray}
&&-B\left( 1-\beta m^{2}\right) \sqrt{G}\partial _{r}I+B\beta \sqrt{G}%
\partial _{r}I\left( g^{rr}\left( \partial _{r}I\right) ^{2}+g^{\xi \xi
}\left( \partial _{\xi }I\right) ^{2}\right)  \notag \\
&&-\frac{iA}{\sqrt{F}}\partial _{t}I-iA\frac{e\hat{A}_{t}}{\sqrt{F}}\left[
1-\beta m^{2}-\left( g^{rr}\left( \partial _{r}I\right) ^{2}+g^{\xi \xi
}\left( \partial _{\xi }I\right) ^{2}\right) \right]  \notag \\
&&+Am\left[ (1-\beta m^{2})-\beta \left( g^{rr}\left( \partial _{r}I\right)
^{2}+g^{\xi \xi }\left( \partial _{\xi }I\right) ^{2}\right) \right] =0,
\label{3.7} \\
&&  \notag \\
&&-A\left( 1-\beta m^{2}\right) \sqrt{G}\partial _{r}I+A\beta \sqrt{G}%
\partial _{r}I\left( g^{rr}\left( \partial _{r}I\right) ^{2}+g^{\xi \xi
}\left( \partial _{\xi }I\right) ^{2}\right)  \notag \\
&&+\frac{iB}{\sqrt{F}}\partial _{t}I+iB\frac{e\hat{A}_{t}}{\sqrt{F}}\left[
1-\beta m^{2}-\left( g^{rr}\left( \partial _{r}I\right) ^{2}+g^{\xi \xi
}\left( \partial _{\xi }I\right) ^{2}\right) \right]  \notag \\
&&+Bm\left[ (1-\beta m^{2})-\beta \left( g^{rr}\left( \partial _{r}I\right)
^{2}+g^{\xi \xi }\left( \partial _{\xi }I\right) ^{2}\right) \right] =0,
\label{3.8} \\
&&  \notag \\
&&A\sqrt{g^{\xi \xi }}\partial _{\xi }I\left[ -\left( 1-\beta m^{2}\right)
+\beta \left( g^{rr}\left( \partial _{r}I\right) ^{2}+g^{\xi \xi }\left(
\partial _{\xi }I\right) ^{2}\right) \right] =0. \label{3.9} 
\end{eqnarray}%
It is hard to directly solve the above system of coupled equations for
action. Indeed, the line element (\ref{1}), equivalently (\ref{5}), is
stationary and admit a Killing vector field $\partial _{t}$, so we decompose
the action $I$ as 
\begin{equation}
I=-\left( \omega -j\Omega \right) t+W\left( r,\xi \right) ,  \label{3.11}
\end{equation}%
where $\omega $ and $j$ are the energy and angular momentum of the emitted
fermion. From (\ref{3.7}) and (\ref{3.8}) it is easy to see that for
nonzero zero wave function, $\partial _{\xi }I=0,$ which just mean that in
dragging coordinates action is independent of $\xi $. Thus, from now on
without loss of generality we fix $\xi =\xi _{0}$. Inserting $\left( \ref%
{3.11}\right) $ into $\left( \ref{3.7}\right) $ and $\left( \ref{3.8}\right) 
$ we obtained 
\begin{eqnarray}
&&\frac{iA}{\sqrt{F}}\left( \omega -j\Omega \right) -iA\frac{e\hat{A}_{t}}{%
\sqrt{F}}\left[ 1-\beta m^{2}-g^{rr}\left( \partial _{r}W\right) ^{2}\right]
+B\beta g^{rr}\sqrt{G}\left( \partial _{r}W\right) ^{3}  \notag \\
&&-B\left( 1-\beta m^{2}\right) \sqrt{G}\partial _{r}W+Bm\left[ (1-\beta
m^{2})-\beta g^{rr}\left( \partial _{r}W\right) ^{2}\right] =0,  \label{3.13}
\end{eqnarray}%
\begin{eqnarray}
&&-\frac{iB}{\sqrt{F}}\left( \omega -j\Omega \right) +iB\frac{e\hat{A}_{t}}{%
\sqrt{F}}\left[ 1-\beta m^{2}-g^{rr}\left( \partial _{r}W\right) ^{2}\right]
+A\beta g^{rr}\sqrt{G}\left( \partial _{r}W\right) ^{3}  \notag \\
&&-A\left( 1-\beta m^{2}\right) \sqrt{G}\partial _{r}W+Bm\left[ (1-\beta
m^{2})-\beta g^{rr}\left( \partial _{r}W\right) ^{2}\right] =0.  \label{3.14}
\end{eqnarray}%
This is system of homogeneous equations in $A$ and $B$ and have nontrivial
solution, if determinant of the coefficient matrix vanishes, thus we must
have%
\begin{equation}
A_{6}\left( \partial _{r}W\right) ^{6}+A_{4}\left( \partial _{r}W\right)
^{4}+A_{2}\left( \partial _{r}W\right) ^{2}+A_{0}=0,  \label{3.15}
\end{equation}%
where%
\begin{eqnarray*}
A_{6} &=&\beta ^{2}G^{3}F, \\
A_{4} &=&\beta G^{2}F\left( m^{2}\beta -2\right) -\beta ^{2}G^{2}e^{2}\hat{A}%
_{t}^{2}, \\
A_{2} &=&GF\left( 1-\beta m^{2}\right) \left( 1+\beta m^{2}\right) +2\beta e%
\hat{A}_{t}G\left[ -\left( \omega -j\Omega \right) +e\hat{A}_{t}\left(
1-\beta m^{2}\right) \right] , \\
A_{0} &=&-m^{2}F\left( 1-\beta m^{2}\right) ^{2}-\left[ \left( \omega
-j\Omega \right) -e\hat{A}_{t}\left( 1-\beta m^{2}\right) \right] ^{2}.
\end{eqnarray*}%
Solving $\left( \ref{3.15}\right) $ by neglecting higher powers of $\beta $
we get%
\begin{equation}
W_{\pm }=\pm \int \sqrt{\frac{m^{2}F+\left[ \left( \omega -j\Omega \right)
-e\hat{A}_{t}\left( 1-\beta m^{2}\right) \right] ^{2}}{FG}}\left[ 1+\beta 
\frac{m^{2}F+\omega _{0}^{2}-e\hat{A}_{t}\omega _{0}}{F}\right] dr,
\label{3.16}
\end{equation}%
where $\omega _{0}=\omega -j\Omega -e\hat{A}_{t}$ and $+/-$ corresponds to
outgoing/ingoing solutions$.$ The above integral equation has pole at
horizons of the black hole and can be solved by complex contour integration.
Note that (\ref{3.16}) has pole of order $2$ at horizons and thus instead of
using $F\left( r\right) =\left( r-r_{h}\right) F^{^{\prime }}\left(
r_{h}\right) $ and $G\left( r\right) =\left( r-r_{h}\right) G^{^{\prime
}}\left( r_{h}\right) $ by Taylor theorem, we use factor theorem so that $%
F_{1}\left( r\right) =F\left( r\right) /\left( r-r_{h}\right) $ and $%
G_{1}\left( r\right) =G(r)/\left( r-r_{h}\right) $. Solving around the
horizon $r_{h}$ with fixed $\xi =\xi _{0},$ we get 
\begin{equation}
Im\left( W_{\pm }\right) =\pm \pi \frac{\left( \omega -j\Omega
_{h}-eA_{t_{h}}\right) }{\sqrt{F_{1}\left( r_{h}\right) G_{1}\left(
r_{h}\right) }}\left( 1+\beta \Xi \right) ,  \label{3.17}
\end{equation}%
where%
\begin{eqnarray}
\Xi &=&\frac{3}{2}m^{2}+\frac{m^{2}e\hat{A}_{t_{h}}}{2\bar{\omega}_{0}}%
-\left( \frac{2e\left( 2\bar{\omega}_{0}-e\hat{A}_{t_{h}}\right) \hat{A}%
_{t_{h}}^{^{\prime }}+j\left( 3\bar{\omega}_{0}-2e\hat{A}_{t_{h}}\right)
\Omega _{h}^{^{\prime }}}{F_{1}\left( r_{+}\right) }\right)  \notag \\
&&-\frac{1}{2}\frac{\bar{\omega}_{0}\left( \bar{\omega}_{0}-e\hat{A}%
_{t_{h}}\right) }{F_{1}\left( r_{h}\right) }\left( 3\frac{F_{1}^{^{\prime
}}\left( r_{h}\right) }{F_{1}\left( r_{h}\right) }+\frac{G_{1}^{^{\prime
}}\left( r_{h}\right) }{G_{1}\left( r_{h}\right) }\right) .
\label{correction}
\end{eqnarray}%
Here $\hat{A}_{t_{h}}=\hat{A}_{t}\left( r_{h}\right) ,$ $\Omega _{h}=\Omega
\left( r_{h}\right) $, $\bar{\omega}_{0}=\omega -j\Omega _{h}-e\hat{A}%
_{t_{h}}$ and prime denotes derivative with respect to $r.$ The tunneling
probability of the fermions, with the contribution of temporal part is given
as \cite{1410.5075v1}%
\begin{equation}
\Gamma \varpropto \exp \left[ -\left( Im\left( \bar{\omega}_{0}\Delta
t^{out,in}\right) +Im p_{r}dr\right) \right] ,  \label{3.19}
\end{equation}%
where 
\begin{equation*}
Im\left( \bar{\omega}_{0}\Delta t^{out,in}\right) =\frac{\pi \bar{%
\omega}_{0}}{2\kappa _{h}},
\end{equation*}%
with $\kappa _{h}$ is the standard surface gravity of corresponding
stationary axially symmetric black hole at horizon $r_{h}$. Considering
total temporal contribution $Im\left( \bar{\omega}_{0}\Delta t\right)
=\pi \bar{\omega}_{0}/\kappa _{h},$ we get the expression of the tunneling
probability%
\begin{eqnarray}
\Gamma &=&\exp \left[ -\left( Im\left( \bar{\omega}_{0}\Delta
t\right) +2ImW^{out}\right) \right] ,  \label{3.21} \\
\Gamma &=&\exp \left[ -4\pi \frac{\left( \omega -j\Omega
_{+}-eA_{t_{+}}\right) }{\sqrt{F_{1}\left( r_{+}\right) G_{1}\left(
r_{+}\right) }}\left( 1+\frac{\beta}{2}\Xi \right) \right] .  \label{3.22}
\end{eqnarray}%
Thus, the modified Hawking temperature for black hole with line element (\ref%
{1}) reads the value%
\begin{equation}
T=\frac{1}{4\pi }\frac{\sqrt{F_{1}\left( r_{h}\right) G_{1}\left(
r_{h}\right) }}{\left( 1+\frac{\beta}{2}\Xi \right) }=T_{0}\left( 1-\frac{\beta}{2}%
 \Xi \right) ,  \label{MT}
\end{equation}%
where $T_{0}=\sqrt{F_{1}\left( r_{h}\right) G_{1}\left( r_{h}\right) }/4\pi $
is the standard Hawking temperature of (\ref{1}). Note that for positive
temperature $\frac{2}{\beta }>\Xi $ and for $\Xi >0$ the modified
temperature is lower then standard temperature. 


\section{Modified temperature of Kerr-Newman black hole}

In this section, we use the general formula derived for modified Hawking
temperature in last section to find modified temperature of Kerr-Newman
black hole. The Kerr-Newman black hole is stationary axially symmetric black
hole and its line element share the form of (\ref{1}) with $\xi =\theta $ as 
\cite{KN}%
\begin{eqnarray}
ds^{2} &=&-\frac{\Delta -a^{2}\sin ^{2}\theta }{\Sigma }dt^{2}+\frac{\Sigma 
}{\Delta }dr^{2}+\Sigma d\theta ^{2}+\frac{\left( r^{2}+a^{2}\right)
^{2}-\Delta a^{2}\sin ^{2}\theta }{\Sigma }d\phi ^{2}  \notag \\
&&-2a\sin ^{2}\theta \frac{r^{2}+a^{2}-\Delta }{\Sigma }dtd\phi ,
\label{3.1}
\end{eqnarray}%
with the electromagnetic potential%
\begin{equation}
A_{\mu }=\frac{Qr}{\Sigma }dt-\frac{Qra\sin ^{2}\theta }{\Sigma }d\phi ,
\label{3.1b}
\end{equation}%
where $\Sigma =r^{2}+a^{2}\cos ^{2}\theta $, $\Delta
=r^{2}+a^{2}+Q^{2}-2Mr=\left( r-r_{+}\right) \left( r-r_{-}\right) $ and the
parameter $M,$ $Q$ and $a$ denote the mass, electric charge and angular
momentum per unit mass, respectively. The outer and inner horizons are
located at $r_{\pm }=M\pm \sqrt{M^{2}-Q^{2}-a^{2}}$. The angular velocity
for the Kerr-Newman black hole given by 
\begin{equation}
\Omega =-\frac{g_{t\phi }}{g_{\phi \phi }}=\frac{a\left( r^{2}+a^{2}-\Delta
\right) }{K},  \label{3.2}
\end{equation}%
where $K=\left( r^{2}+a^{2}\right) ^{2}-\Delta a^{2}\sin ^{2}\theta .$ Using
dragged coordinate transformation $d\phi =\Omega dt,$ with angular velocity (%
\ref{3.2}), the dragged line element and corresponding electromagnetic
potential of Kerr-Newman black hole takes the form%
\begin{equation}
d\hat{s}^{2}=-\frac{\Sigma \Delta }{K}dt^{2}+\frac{\Sigma }{\Delta }%
dr^{2}+\Sigma d\theta ^{2},
\end{equation}%
and%
\begin{equation}
\hat{A}_{i}=\hat{A}_{t}dt=\frac{Qr\left( r^{2}+a^{2}\right) }{K}dt.
\label{EP}
\end{equation}%
To determine modified temperature at outer horizon $r_{+}$\ the functions $%
F_{1}$ and $G_{1}$ are%
\begin{equation}
F_{1}=\frac{\left( r-r_{-}\right) \Sigma }{K}\text{ and }G_{1}=\frac{\left(
r-r_{-}\right) }{\Sigma }.  \label{fg}
\end{equation}%
Now we are in position to find modified temperature for Kerr-Newman black
hole using the formula (\ref{MT}) with correction terms given by (\ref%
{correction}). Using angular velocity (\ref{3.2}), electromagnetic potential
(\ref{EP}) and functions (\ref{fg}) into (\ref{correction}) we get%
\begin{eqnarray}
\Xi _{_{KN}} &=&\frac{3}{2}m^{2}+\frac{m^{2}e\hat{A}_{t+}}{2\bar{\omega}_{0}}%
+\frac{2eQ\left( 2\bar{\omega}_{0}-e\hat{A}_{t+}\right) }{\Sigma \left(
r_{+}\right) \left( r_{+}-r_{-}\right) }\left[ r_{+}^{2}-a^{2}-\frac{%
r_{+}\left( r_{+}-r_{-}\right) }{r_{+}^{2}+a^{2}}a^{2}\sin ^{2}\theta _{0}%
\right]  \notag \\
&&+\frac{\left( 3\bar{\omega}_{0}-2e\hat{A}_{t+}\right) }{\Sigma \left(
r_{+}\right) \left( r_{+}-r_{-}\right) }\left[ j\Omega _{+}\left\{
4r_{+}\left( r_{+}^{2}+a^{2}\right) -\left( r_{+}-r_{-}\right) a^{2}\sin
^{2}\theta _{0}\right\} \right.  \notag \\
&&\left. -ja\left( r_{+}+r_{-}\right) \right] +\frac{\bar{\omega}_{0}\left( 
\bar{\omega}_{0}-e\hat{A}_{t+}\right) }{2\Sigma \left( r_{+}\right) \left(
r_{+}-r_{-}\right) }\left[ 12r_{+}\left( r_{+}^{2}+a^{2}\right) -\frac{%
4(r_{+}^{2}+a^{2})^{2}}{r_{+}-r_{-}}\right.  \notag \\
&&\left. -3\left( r_{+}-r_{-}\right) a^{2}\sin ^{2}\theta _{0}-\frac{%
4r_{+}(r_{+}^{2}+a^{2})^{2}}{\Sigma \left( r_{+}\right) }\right],
\end{eqnarray}%
with 
\begin{equation}
\bar{\omega}_{0}=\omega -j\Omega _{+}-e\hat{A}_{t+},\text{ \ \ }\Omega _{+}=%
\frac{a}{r_{+}^{2}+a^{2}},\text{ \ }\hat{A}_{t+}=\frac{Qr_{+}}{%
r_{+}^{2}+a^{2}}.
\end{equation}%
Thus, the modified Hawking temperature for Kerr-Newman black hole is 
\begin{equation}
T=\frac{1}{4\pi }\frac{r_{+}-r_{-}}{r_{+}^{2}+a^{2}}\left( 1+\frac{1}{2}%
\beta \Xi _{_{KN}}\right) ^{-1}=T_{0}\left( 1-\frac{1}{2}\beta \Xi
_{_{KN}}\right) ,
\end{equation}%
where $T_{0}$ is standard Hawking temperature for Kerr-Newman black hole.
When $a=0$ and $q=0$ the modified temperature reduces to the Reissner-Nordstr\"{o}m and Kerr black holes, respectively. Due to $\theta _{0}$ in the
correction term, the modified
temperature depends on angle $\theta $. Their claim is not in agreement with
zeroth law of thermodynamics. So for
constant temperature everywhere on the horizons we can set $\theta _{0}=0,$
and in this case correction terms $\Xi _{_{KN}}$ reduces to%
\begin{eqnarray}
\Xi _{_{KN}} &=&\frac{3}{2}m^{2}+\frac{m^{2}e\hat{A}_{t+}}{2\bar{\omega}_{0}}%
+\frac{2eQ\left( 2\bar{\omega}_{0}-e\hat{A}_{t+}\right) \left(
r_{+}^{2}-a^{2}\right) }{\left( r_{+}-r_{-}\right) \left(
r_{+}^{2}+a^{2}\right) }  \notag \\
&&+\frac{aj\left( 3\bar{\omega}_{0}-2e\hat{A}_{t+}\right) }{\left(
r_{+}-r_{-}\right) \left( r_{+}^{2}+a^{2}\right) }\left[ 4r_{+}\left(
r_{+}^{2}+a^{2}\right) -\left( r_{+}+r_{-}\right) \right]  \notag \\
&&+\frac{2\bar{\omega}_{0}\left( \bar{\omega}_{0}-e\hat{A}_{t+}\right) }{%
\left( r_{+}-r_{-}\right) }\left[ 2r_{+}-\frac{(r_{+}^{2}+a^{2})}{r_{+}-r_{-}%
}\right].
\end{eqnarray}%
Using $r_{\pm }=M\pm \sqrt{M^{2}-Q^{2}-a^{2}}$ it can be easily shown that $%
\Xi _{KN}>0$ thus the modified temperature is lower than that of standard
temperature and for positive temperature it must be in the limit $\Xi
_{_{KN}}<\frac{2}{\beta }.$ Further, if we ignore the quantum gravity
effects we will get the standard temperature for Kerr-Newmen black holes.

\section{Modified temperature for EMDA black hole}

In this section we give the modified temperature for EMDA black hole. The
EMDA black hole is stationary axially symmetric solution of the
Einstein-Maxwell Dilaton-Axion field equations.The line element of EMDA
black hole of mass $M,$ angular momentum per unit mass $a$ and dilatonic
perimeter $b$ is given by \cite{1204.4319v2(26)}%
\begin{eqnarray}
ds^{2} &=&-\frac{\Delta -a^{2}\sin ^{2}\theta }{\Sigma }dt^{2}+\frac{\Sigma 
}{\Delta }dr^{2}+\Sigma d\theta ^{2}-2\frac{a\sin ^{2}\theta \left(
r^{2}+2br+a^{2}-\Delta \right) }{\Sigma }dtd\phi  \notag \\
&&+\frac{\left( r^{2}+2br+a^{2}\right) ^{2}-\Delta a^{2}\sin ^{2}\theta }{%
\Sigma }\sin ^{2}\theta d\phi ^{2}.
\end{eqnarray}%
The corresponding electromagnetic vector potential is%
\begin{equation}
A_{\mu }=\frac{Qr}{\Sigma }dt-\frac{Qra\sin ^{2}\theta }{\Sigma }d\phi ,\label{pote}
\end{equation}%
where $\Sigma =r^{2}+2br+a^{2}\cos ^{2}\theta ,$ $\Delta
=r^{2}-2Mr+a^{2}=\left( r-r_{+}\right) \left( r-r_{-}\right) $ and $r_{\pm
}=M\pm \sqrt{M^{2}-a^{2}}$ are location of the outer and inner horizons. The
ADM mass $M_{A},$ charge $Q$ and the angular momentum $J$ are related with
diatonic parameter as 
\begin{equation*}
M_{A}=M+b,\text{ \ \ }Q^{2}=2b(M+b),\text{ \ \ }J=\left( M+b\right) a.
\end{equation*}%
The EMDA black hole is generalization of the Kerr and the Garfinkle-Horowitz-Strominger dilatonic (GHSD) black holes, with parameter $%
b $ and $a,$ respectively$.$ The angular velocity for this black hole is
given as%
\begin{equation}
\Omega =\frac{a\left( r^{2}+2br+a^{2}-\Delta \right) }{K},  \label{AV}
\end{equation}%
with $K=\left( r^{2}+2br+a^{2}\right) ^{2}-\Delta a^{2}\sin ^{2}\theta .$
With this angular velocity we obtained the transformed dragged line element
as 
\begin{equation}
d\hat{s}^{2}=-\frac{\Delta \Sigma }{K}dt^{2}+\frac{\Sigma }{\Delta }%
dr^{2}+\Sigma d\theta ^{2},
\end{equation}%
and the electromagnetic potential%
\begin{equation}
A_{i}=\hat{A}_{t}dt=\frac{Qr\left( r^{2}+2br+a^{2}\right) }{K}dt.
\label{EP EDMA}
\end{equation}%
Using the factor theorem for outer horizon we can define the functions as 
\begin{equation}
F_{1}=\frac{\left( r-r_{-}\right) \Sigma }{K}\text{ and }G_{1}=\frac{\left(
r-r_{-}\right) }{\Sigma }. \label{fg EDMA}
\end{equation}%
Using angular velocity (\ref{AV}), electromagnetic potential (\ref{EP EDMA}) and functions (\ref{fg EDMA}) we have
\begin{eqnarray}
\Xi _{EDMA} &=&\frac{3}{2}m^{2}+\frac{m^{2}e\hat{A}_{t+}}{2\bar{\omega}_{0}}+%
\frac{2eQ\left( 2\bar{\omega}_{0}-e\hat{A}_{t+}\right) \left(
r_{+}^{2}-a^{2}\right) }{\Sigma \left( r_{+}\right) \left(
r_{+}-r_{-}\right) }  \notag \\
&&+\frac{aj\left( 3\bar{\omega}_{0}-2e\hat{A}_{t+}\right) }{\left(
r_{+}-r_{-}\right) }\frac{\left[ 4\left( r_{+}+b\right) -\left(
r_{+}+2b+r_{-}\right) \right] }{\left( r_{+}^{2}+2br_{+}+a^{2}\right) } \notag\\
&&+\frac{\bar{\omega}_{0}\left( \bar{\omega}_{0}-e\hat{A}_{t+}\right) }{%
\left( r_{+}-r_{-}\right) }\left[ 4\left( r_{+}+b\right) -\frac{%
2(r_{+}^{2}+2br_{+}+a^{2})}{r_{+}-r_{-}}\right] ,
\end{eqnarray}%
with 
\begin{equation}
\bar{\omega}_{0}=\omega -j\Omega _{+}-e\hat{A}_{t+},\text{ \ \ }\Omega _{+}=%
\frac{a}{r_{+}^{2}+2br_{+}+a^{2}},\text{ \ \ }A_{t+}=\frac{Qr_{+}}{%
r_{+}^{2}+2br_{+}+a^{2}}.
\end{equation}%
Thus modified Hawking temperature for EDMA black hole is given as%
\begin{equation}
T=\frac{1}{4\pi }\left( \frac{r_{+}^{2}+2br_{+}+a^{2}}{r_{+}-r_{-}}\right) 
\frac{1}{\left( 1+\frac{1}{2}\Xi _{EDMA}\right) }=T_{0}\left( 1-\frac{1}{2}%
\beta \Xi _{EDMA}\right).   \label{3.23}
\end{equation}%
When $\ a=0,$ we will get modified temperature for GHSD black holes.
Further, if $\Xi _{EDMA}>0,$ the modified temperature is less than the
standard temperature.

\section{Modified temperature for Kaluza-Klein dilaton black hole}

The Kaluza-Klein black hole is an exact solution of the dilatonic action
with coupling constant $\alpha =3.$ It is derived by a dimensional reduction
of the boosted five dimensional Kerr solution to four dimensions. The line
element of Kaluza-Klein dilaton black hole is given by \cite{bh1,bh2}%
\begin{eqnarray}
ds^{2} &=&-\frac{\Delta -a^{2}\sin ^{2}\theta }{\Pi \Sigma }dt^{2}+\frac{\Pi
\Sigma }{\Delta }dr^{2}+\Pi \Sigma d\theta ^{2}-2\frac{aZ\sin ^{2}\theta }{%
\Pi \sqrt{1-\nu ^{2}}}dtd\phi  \notag \\
&&+\left[ \Pi \left( r^{2}+a^{2}\right) +\frac{Z}{\Pi }a^{2}\sin ^{2}\theta %
\right] \sin ^{2}\theta d\phi ^{2},
\end{eqnarray}%
with the electromagnetic potential and the dilaton field 
\begin{eqnarray}
A_{\mu } &=&\frac{\nu }{2\left( 1-\nu ^{2}\right) }\frac{Z}{\Pi ^{2}}dt-%
\frac{\nu a\sin ^{2}\theta }{2\sqrt{1-\nu ^{2}}}\frac{Z}{\Pi ^{2}}d\phi , \\
\Phi &=&-\frac{\sqrt{3}\ln \Pi }{2},
\end{eqnarray}%
where%
\begin{eqnarray}
Z &=&\frac{2\mu r}{\Sigma },\text{ }\Pi =\sqrt{1+\frac{\nu ^{2}Z}{1-\nu ^{2}}%
}, \\
\Sigma &=&r^{2}+a^{2}\cos ^{2}\theta ,\text{ }\Delta =r^{2}-2Mr+a^{2},
\end{eqnarray}%
and $M,$ $a$ and $\nu $ are the mass parameter specific angular momentum and
boosted velocity, respectively. The horizons are located at $r_{\pm }=M\pm 
\sqrt{\mu ^{2}-a^{2}}.$ The physical mass $M_{p},$ the charge $Q$ and the
angular momentum $J$ can be related with mass parameter and boosted velocity
and specific angular momentum as%
\begin{equation}
M_{p}=M\left[ 1+\frac{\nu ^{2}}{2\left( 1-\nu ^{2}\right) }\right] ,\text{ \
\ }Q=\frac{M\nu }{1-\nu ^{2}},\text{ \ \ }J=\frac{Ma}{\sqrt{1-\nu ^{2}}}.
\end{equation}%
The angular velocity for this black hole 
\begin{equation}
\Omega =\frac{aZ\sqrt{1-\nu ^{2}}}{\left( r^{2}+a^{2}\right) Z+\left( 1-\nu
^{2}\right) \Delta }.  \label{AVKK}
\end{equation}%
Using this angular velocity the dragged line element becomes%
\begin{equation}
d\hat{s}^{2}=-\frac{\Pi \Sigma \Delta \left( 1-v^{2}\right) }{K}dt^{2}+\frac{%
\Pi \Sigma }{\Delta }dr^{2}+\Pi \Sigma d\theta ^{2},
\end{equation}%
with $K=\left( r^{2}+a^{2}\right) ^{2}-\Delta \left( a^{2}\sin ^{2}\theta
+\nu ^{2}\Sigma \right) $. The dragged electromagnetic potential is 
\begin{equation}
\hat{A}_{i}=\hat{A}_{t}dt=\frac{Qr\left( r^{2}+2br+a^{2}\right) }{K}dt.
\label{KKEP}
\end{equation}%
For modified temperature at horizon $r_{+}$ the functions are 
\begin{equation}
F_{1}=\frac{\Pi \Sigma \left( 1-\nu ^{2}\right) \left( r-r_{-}\right) }{K}%
\text{ and }G_{1}=\frac{\left( r-r_{-}\right) }{\Pi \Sigma }. \label{KKfg}
\end{equation}%
Using angular velocity (\ref{AVKK}), electromagnetic potential (\ref{KKEP})
and the functions (\ref{KKfg}) we get%
\begin{eqnarray}
\Xi _{KK} &=&\frac{3}{2}m^{2}+\frac{m^{2}e\hat{A}_{t_{+}}}{2\omega _{0}}-2%
\frac{e\hat{A}_{t_{+}}\left( 2\omega _{0}-e\hat{A}_{t_{+}}\right) }{%
F_{1}\left( r_{\alpha }\right) }\left[ \frac{2r}{r^{2}+a^{2}}+\frac{1}{r}-%
\frac{2r}{\Sigma }-\right. \notag\\
&&\left. \frac{2\Pi _{+}^{^{\prime }}\left( r_{+}^{2}+a^{2}\right)
+4r_{+}\Pi _{+}}{\Pi _{+}\left( r_{+}^{2}+a^{2}\right) }\right] -\frac{1}{2}%
\frac{\omega _{0}\left( \omega _{0}-e\hat{A}_{t_{+}}\right) }{\Pi _{+}\left(
1-\nu ^{2}\right) \left( r_{+}-r_{-}\right) \left( r_{+}^{2}+a^{2}\right) }%
\times \notag\\
&&\left[ \frac{4\left( r_{+}^{2}+a^{2}\right) ^{2}}{r_{+}-r_{-}}+\frac{2m\nu
^{2}\left( a^{2}-r_{+}^{2}\right) }{\left[ 1+\nu ^{2}\left( Z_{+}-1\right) %
\right] }-\left( r_{+}^{2}+a^{2}\right) \left\{ 8r_{+}+6\left( r_{+}-\mu
\right) \nu ^{2}\right\} \right] \notag\\
&&+\frac{aj\left( 3\omega _{0}-2e\hat{A}_{t_{+}}\right) }{\sqrt{1-\nu ^{2}}}%
\left[ \frac{2r_{+}Z_{+}+\left( 1-\nu ^{2}\right) \left( r_{+}-r_{-}\right) 
}{Z_{+}\Pi _{+}\left( r_{+}-r_{-}\right) \left( r_{+}^{2}+a^{2}\right) }%
\right] ,
\end{eqnarray}%
where%
\begin{equation}
\bar{\omega}_{0}=\omega -j\Omega _{+}-e\hat{A}_{t+},\text{ \ \ }\Omega _{+}=%
\frac{a\sqrt{1-\nu ^{2}}}{r^{2}+a^{2}}, \text{ \ \ }A_{t+}=\frac{Qr_{+}}{%
r_{+}^{2}+2br_{+}+a^{2}}.
\end{equation}

\section{Modified temperature for charged rotating black strings}

The line element of charged rotating black string is stationary and axially
symmetric which admit three Killing vectors, $\partial _{t},$ $\partial
_{\phi },$ $\partial _{z}$. Thus the modified temperature for black string
can be obtained from general formula given in Section $2$ with $\xi =z.$ The
line element of charged rotating black strings can be written as \cite{bstring2} 
\begin{equation}
ds^{2}=-\Delta \left( \gamma dt-\frac{\delta }{\alpha ^{2}}d\phi \right)
^{2}+r^{2}\left( \gamma d\phi -\delta dt\right) ^{2}+\frac{dr^{2}}{\Delta }%
+\alpha ^{2}r^{2}dz^{2},  \label{BS}
\end{equation}%
with%
\begin{eqnarray}
\Delta &=&\alpha ^{2}r^{2}-\frac{b}{\alpha r}+\frac{c^{2}}{\alpha ^{2}r^{2}},%
\text{ \ }b=4M\left( 1-\frac{3a^{2}\alpha ^{2}}{2}\right) ,\text{ \ }%
c^{2}=4Q^{2}\left( \frac{2-3a^{2}\alpha ^{2}}{2-a^{2}\alpha ^{2}}\right) , 
 \\
\gamma &=&\sqrt[\backslash ]{\frac{2-a^{2}\alpha ^{2}}{2-2a^{2}\alpha ^{2}},}%
\text{ \ \ }\delta =\frac{a\alpha ^{2}}{\sqrt{1-\frac{3}{2}a^{2}\alpha ^{2}}}%
.  \label{BS1}
\end{eqnarray}%
where $M$ is mass, $a$ is the angular momentum per unit mass, $\alpha ^{2}=-%
\frac{\Lambda }{3},$ $\Lambda $ is the negative cosmological constant. The
parameters $a$ and $\alpha ^{2}$ are related with angular momentum $J$ and
the mass of the black hole $M$ as%
\begin{eqnarray}
a^{2}\alpha ^{2} &=&1-\frac{\sqrt{M^{2}-\frac{8J^{2}\alpha ^{2}}{9}}}{M}, \\
J &=&\frac{3}{2}Ma\sqrt{1-\frac{a^{2}\alpha ^{2}}{2}.}
\end{eqnarray}%
The corresponding electromagnetic potential is given by%
\begin{equation}
A_{\mu }=\frac{2Q}{\alpha r}dt-\frac{2\delta Q}{\alpha ^{3}r\gamma }d\phi.
\end{equation}%
The angular velocity for charged rotating black string is%
\begin{equation}
\Omega =\frac{-\Delta \gamma \delta \alpha ^{2}+r^{2}\gamma \delta \alpha
^{4}}{-\Delta \delta ^{2}+r^{2}\gamma ^{2}\alpha ^{4}}.  \label{BSAV}
\end{equation}%
With this angular velocity after transformation we get dragged metric \cite{BString} 
\begin{eqnarray}
d\hat{s}^{2} &=&-\frac{\Delta r^{2}\alpha ^{4}}{-\Delta \delta
^{2}+r^{2}\gamma ^{2}\alpha ^{4}}dt^{2}+\frac{dr^{2}}{\Delta }+\alpha
^{2}r^{2}dz^{2},  \notag \\
d\hat{s}^{2} &\equiv &-Fdt^{2}+\frac{dr^{2}}{G}+H^{2}d\xi ^{2}.  \label{BS2}
\end{eqnarray}%
The corresponding electromagnetic vector potential is%
\begin{equation}
\hat{A}_{i}=\hat{A}_{t}dt=\left( \frac{2Qr\alpha ^{3}}{-\Delta \delta
^{2}+r^{2}\gamma ^{2}\alpha ^{4}}\right) dt.  \label{BSEP}
\end{equation}%
For modified temperature at horizon of black string $r_{+}$ we have 
\begin{equation}
F_{1}=\frac{\alpha ^{2}\left( \alpha ^{4}r^{3}+\alpha ^{4}r_{+}r^{2}+\alpha
^{4}r_{+}^{2}r-\frac{c^{2}}{r_{+}}\right) }{-\Delta \delta ^{2}+r^{2}\gamma
^{2}\alpha ^{4}},\text{ \ }G_{1}=\alpha ^{2}r+\alpha ^{2}r_{+}+\frac{\alpha
^{2}r_{+}^{2}}{r}-\frac{c^{2}}{\alpha ^{2}r_{+}r^{2}}. \label{BSfg}
\end{equation}%
With angular velocity (\ref{BSAV}), (\ref{BSEP}) and (\ref{BSfg})%
\begin{eqnarray}
\Xi _{BS} &=&\frac{3}{2}m^{2}+\frac{m^{2}e\hat{A}_{t_{+}}}{2\bar{\omega}_{0}}%
+\frac{4eQ\left( 2\bar{\omega}_{0}-e\hat{A}_{t_{+}}\right) }{\alpha
r_{+}^{2}\Delta ^{^{\prime }}\left( r_{+}\right) }\left( 1-\frac{\Delta
^{^{\prime }}\left( r_{+}\right) \delta ^{2}}{r_{+}\gamma ^{2}\alpha ^{4}}%
\right) +\frac{j\left( 3\bar{\omega}_{0}-2e\hat{A}_{t_{+}}\right) \Omega _{+}%
}{\alpha ^{2}r_{+}^{2}}  \notag \\
&&-\frac{1}{2}\frac{\bar{\omega}_{0}\left( \bar{\omega}_{0}-e\hat{A}%
_{t_{+}}\right) \gamma ^{2}}{\Delta ^{^{\prime }}\left( r_{+}\right) }\left[
3\left( \frac{6\alpha ^{2}}{\Delta ^{^{\prime }}\left( r_{+}\right) }+\frac{%
\delta ^{2}}{r_{+}^{2}\gamma ^{2}\alpha ^{4}}\Delta ^{^{\prime }}\left(
r_{+}\right) -\frac{2}{r_{+}}\right) -\frac{\alpha ^{2}r_{+}^{4}}{2c^{2}}%
\Delta ^{^{\prime }}\left( r_{+}\right) \right],
\end{eqnarray}%
where
\begin{equation}
\bar{\omega}_{0}=\omega -j\Omega _{+}-e\hat{A}_{t+},\text{ \ \ }\Omega _{+}=%
\frac{\delta }{\gamma },\text{ \ }\hat{A}_{t_{+}}=\frac{2Q}{\alpha \gamma
^{2}r_{+}},\text{ \ }\Delta ^{^{\prime }}\left( r_{+}\right) =\frac{3\alpha
^{4}r_{+}^{4}-c^{2}}{\alpha ^{2}r_{+}^{3}}.
\end{equation}
For $a=0$, the modified temperatures for charged non-rotating black string and $Q=0$ for uncharged rotating black string are successfully recovered \cite{1404.6375v1}.

\section{Conclusion}
In this paper, we first developed a general method to study the tunneling of charged fermions from the stationary axially symmetric black holes with GUP. The important results are given as follows:
\begin{itemize}
    \item To this end, we modified the Dirac equation using the GUP and solve it using the corresponding curved spacetime via the semiclassical method of WKB and Hamilton-Jacobi approach.
    
     \item After we obtained the corrected tunneling rate of the fermions from the curved spacetime, we showed the modified Hawking temperature for the most general case, then using this method, we gave same examples how to calculate Hawking temperatures of Kerr-Newman black hole, Einstein-Maxwell-Dilaton-Axion (EMDA) black hole, Kaluza-Klein dilaton black hole, and then, charged rotating black string.
     
      \item The corrected thermal spectrum was shown that it is not purely thermal. We noted that the effect of the GUP causes the the black hole's remnant.
      
       \item Moreover, the modified Hawking temperature of the black holes is lower than the standard Hawking temperature.
       \item The remnant of the black hole's radiation increases, when the black hole size is close to the Planck scale, because of the effect of the quantum gravity.

       \item Due to this remnant, the black hole is prevented from evaporation, and its information and singularity are enclosed in the event horizon.
       
       \end{itemize}

\acknowledgments
This work is supported by Comisi\'on Nacional
de Ciencias y Tecnolog\'ia of Chile through FONDECYT Grant N$^{\textup{o}}$ 3170035 (A. \"{O}.).

\end{document}